\documentclass[twocolumn,showpacs,preprintnumbers,amsmath,amssymb]{revtex4}

\usepackage{graphicx}
\usepackage{dcolumn}
\usepackage{bm}

\usepackage{amsmath,amssymb,amsfonts, mathpazo}
\usepackage{epsfig, subfigure}
\usepackage{url}

\newtheorem{thm}{Theorem}

\newcommand{\ms}[1]{\mbox{\scriptsize{#1}}}
\newcommand{\define}{\triangleq}

\newcommand{\kron}{\otimes}   
\newcommand{\trace}[1]{\mbox{Tr}\left[#1\right]}   

\newcommand{\ket}[1]{|#1\rangle} 
\newcommand{\bra}[1]{\langle#1|} 
\newcommand{\braket}[2]{\langle#1|#2\rangle} 

\newcommand{\magnitude}[1]{\left|{#1}\right|}  
\newcommand{\set}[1]{\{#1\}} 

\newcommand{\Prob}{\mbox{Pr}}

\newcommand{\NOT}{\mathsf{NOT}}   
\newcommand{\CNOT}{\mathsf{CNOT}}   

\newcommand{\eye}{I}   
\newcommand{\zero}{\mathbf{0}}    

\newcommand{\info}{\mathbf{A}}    
\newcommand{\code}{\mathbf{C}}    
\newcommand{\encode}{\mathbf{E}}    
\newcommand{\decode}{\mathbf{D}}    
\newcommand{\noise}{\mathbf{N}}    

\newcommand{\Info}{\mathcal{H}_A}    
\newcommand{\Ancilla}{\mathcal{H}_B}    
\newcommand{\Code}{\mathcal{H}_C}    
\newcommand{\CodeSub}{\mathcal{H}_{C_0}}    
\newcommand{\CodeSubK}{\mathcal{H}_{C_k}}    
\newcommand{\CodeSubL}{\mathcal{H}_{C_l}}    
\newcommand{\Envi}{\mathcal{H}_N}    
\newcommand{\Encode}{E}    
\newcommand{\Decode}{D}    
\newcommand{\NoiseSet}{\mathfrak{N}}    
\newcommand{\NoiseOp}{N}    
\newcommand{\NoiseSpace}{\mathcal{N}}    
\newcommand{\ProjectOp}{P}    
\newcommand{\eyeOp}{I}   
\newcommand{\PauliX}{\sigma_x}   
\newcommand{\PauliY}{\sigma_y}   
\newcommand{\PauliZ}{\sigma_z}   

\begin{document}


\title{An Introduction to Error-Correcting Codes: From Classical to Quantum}

\author{ Hsun-Hsien~Chang}
    \affiliation{Carnegie Mellon University.}
    \email{hsunhsien@cmu.edu}

\date{\today}

\begin{abstract}
This report surveys quantum error-correcting codes. As Preskill
claimed in~\cite{Preskill1998-ReliableQC}, 21st century would be the
golden age of quantum error correction. Quantum channels behave
differently from classical channels, so researchers face
difficulties in developing robust quantum codes. Fortunately, the
classical error control methods have been well developed. If we can
learn many lessons from classical coding theory, we can expedite the
development of quantum codes. Scientists have discovered that
quantum error correction shares many concepts with classical
counterpart. Both quantum and classical coding schemes add
redundancy to information to protect against noises. They also have
similar conditions for error detectability and correctability.

\end{abstract}

\maketitle

\tableofcontents

\vspace{1in}
\section{Introduction} \label{sec:Intro}

\begin{figure*}[t]
    \centering
    \setlength{\unitlength}{1.4cm}
    \begin{picture}(7.4,2.5)
        \thicklines
        \put(0.4,0.6){information}
        \put(0.7,0.3){source}
        \put(2,0.5){\vector(1,0){1}}
        \put(3,0){\framebox(2,1){\textbf{channel}}}
        \put(4,1.5){\vector(0,-1){0.5}}
        \put(3,1.5){\makebox(2,0.5){noise}}
        \put(5,0.5){\vector(1,0){1}}
        \put(6.0,0.6){information}
        \put(6.5,0.3){user}
    \end{picture}
    \caption{Block diagram of a communication system. A source
         transmits information to a user through a channel.
         The communication channel is usually imperfect so that
         the information is prone to errors during transmission.}
    \label{Fig:ChannelModel}
    \end{figure*}
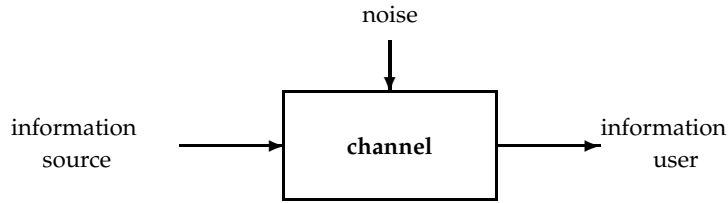

In the age of information technology, error-correcting codes are
widely used in communication systems and data storage systems. Both
types of systems share the same model, as shown in
Fig~\ref{Fig:ChannelModel}. A source transmits information to a user
through a channel. The communication channel, unfortunately, is
usually imperfect; i.e., the information might be corrupted by
noises during transmission. To immunize information to noises, the
sender adds redundancy within the information and follows an
invertible encoding process to mix the redundancy and information.
When the receiver obtains this mixture, it checks where errors are,
corrects the errors as possible, and finally removes redundancy
added by the sender. The above scheme of encoding and decoding is
referred as \textbf{error-correcting codes}.

Coding theory was initiated by two seminal papers:
\begin{enumerate}
\item In 1948, Shannon wrote a detailed treatise on the mathematics
      behind communication \cite{Shannon1948-InfoTheory}.
\item In 1950, Hamming, motivated by the task of correcting a small
      number of errors on magnetic storage media, wrote the first
      paper introducing error-correcting codes
      \cite{Hamming1950-ECC}.
\end{enumerate}
The research area in coding theory has been prosperously progressing
and the theory is well developed. Since there are a tremendous
number of textbooks for coding theory, this report points out only
\emph{basic} principles of error-correcting codes. More details can
be found in standard
textbooks~\cite{Blahut1983-ECC,MacwilliamsSloane1977-ECC}. Other
textbooks written by McEliece~\cite{Mceliece2002-InfoThCoding} and
by Lin and Costello~\cite{LinCostello2004-ECC} cover more up-to-date
coding schemes such as turbo codes and low-density parity-check
codes.

Parallel to the fast progress of coding theory in communications,
the trend in electronics is to shrink the sizes of computing units
and hence to integrate computation, communication and storage
components into single chips. We believe that this trend finally
will force us to design computation, communication and storage
devices in the \emph{quantum} world. To recover the corrupted
information at the output of quantum communication channels, we have
to study \emph{quantum error-correcting codes}, which is the goal of
this report. Although quantum error correction is a new research
area, its foundation is based on classical error correction. We
start the discussion of quantum error correction by introducing the
fundamental principles learned from classical error correction,
described in Section~\ref{sec:Principle}.  We then move on to
quantum error correction in Section~\ref{sec:QuantECC}. Although we
have the capability to protect quantum information against noises in
channels, the quantum encoding/decoding procedure itself is
vulnerable to errors as well. To protect information against errors
during encoding and decoding, Section~\ref{sec:FaultTole} addresses
fault-tolerance in quantum computation. Finally,
Section~\ref{sec:FaultTole} concludes this report.


\section{Principles of Classical Error Correction} \label{sec:Principle}

\begin{figure}[t]
    \centering
    \setlength{\unitlength}{3cm}
    \begin{picture}(3,1)
        \thicklines
        \put(0.8,0.65){1}
        \put(0.8,0.15){0}
        \put(1,0.2){\vector(1,0){1}}
        \put(1,0.2){\vector(2,1){0.9}}
        \put(1,0.7){\vector(1,0){1}}
        \put(1,0.7){\vector(2,-1){0.9}}
        \put(2.1,0.65){1}
        \put(2.1,0.15){0}
        \put(1.3,0.8){$1-p$}
        \put(1.3,0.05){$1-p$}
        \put(1.15,0.52){$p$}
        \put(1.15,0.33){$p$}
    \end{picture}
        \caption{Binary Symmetric Channel.}
        \label{Fig:BSChannel}
    \end{figure}
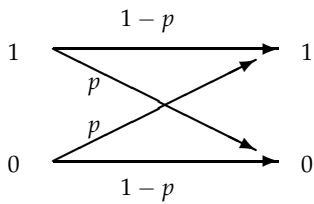

In modern information systems, the unit of information is
\textbf{bit}. A bit takes one of two values: 0 or 1. As mentioned in
Section~\ref{sec:Intro}, a bit transmitted through a channel is
prone to errors. At the end of a channel, an information bit 0/1 may
remain as 0/1 or flip to 1/0. A simple model of noisy channels is
\emph{Binary Symmetric Channel} with parameter $p$ which flips each
transmitted bit with probability $p$ independent of all other
events. This effect is sketched in Fig~\ref{Fig:BSChannel}. Using
this specific channel is enough to illustrate the basic ideas of
error correction.

Before proceeding to the detailed discussions, we first introduce
the notations. Let the set $\info$ consist of information bit
strings. An encoding operator $\encode$ maps $\info$ into a space
$\code$ called \textbf{code}. The elements in the code $\code$ are
called \textbf{codewords}. In the channel, a set of noise operators
$\noise=\set{N_0=\eye,N_1,N_2,\cdots}$ corrupts the codewords. In
$\noise$, $\eye$ is an identity which does nothing wrong to
codewords. All the possible corrupted codewords are collected in a
set $\code'$. A decoding operator $\decode$ at receiver recovers the
received codewords in $\code'$ back to the information strings of
bits. We in Sections~\ref{ssec:ClassicalECC}
and~\ref{ssec:ClassicalErrorDetCorr} use classical examples to
illustrate the ideas of error correction.

\subsection{Learning from Classical Error-Correcting Codes}\label{ssec:ClassicalECC}

The design of error-correcting codes is based on the concept of
\emph{adding redundancy}. This concept happens in our oral
communications too. When people have a discussion, they usually
convey a viewpoint several times or state it in other words. This is
equivalent to repeating the information or changing the wordings of
the information. Same ideas apply to error-correcting codes. We can
encode a bit by repeating it or by using another longer string of
bits to represent it. We now use two examples to illustrate these
two types of ideas.

\underline{\textbf{The Repetition Code:}} Encoding a bit by
repeating it several times is called \textbf{repetition code}. In
the case of triplicating the information bit, we have
$\info=\set{0,1}$ and code $\code=\set{000,111}$. The received
codewords can be decoded by majority voting $\decode_{\ms{mv}}$:
decide 0 if the majority of the codeword is 0, otherwise decide 1.

In the repetition code, we define $N_i$ as a noise operator that has
probability $p$ to flip the $i$th bit. If there is only one noise
operator $N_1$ corrupting the codewords, i.e.,
$\noise=\set{N_0=\eye,N_1}$, the set $\code'$ of all possibly
corrupted codewords is
\begin{equation}
    \code'=\set{000,100,011,111}.
\end{equation}
Based on majority voting, the received codewords
000,100 are being mapped to 0 and 011, 111 to 1. This results in a
perfect recovery of information.

We further add another noise operator $N_2$ affecting the second
bit, so $\noise=\set{N_0,N_1,N_2}$. The corrupted code now becomes
\begin{equation}
    \code'=\set{000,010,100,110, 001,011,101,111}.
\end{equation}
Unfortunately, we this time cannot correct all the
errors using majority voting, because 110 and 001 are misclassified
as 1 and 0, respectively. However, we can design another error
control scheme $\decode_{\ms{3rd}}$: decide 0 if the third bit of
the received codeword is 0, otherwise decide 1. Apparently,
$\decode_{\ms{3rd}}$ is better than $\decode_{\ms{mv}}$ because
$\decode_{\ms{3rd}}$ is able to recover the information bit without
any decoding failures.

How good $\decode_{\ms{3rd}}$ is over $\decode_{\ms{mv}}$ is the
next topic we want to address. In general, we use \emph{probability
of failure $\Pr(\epsilon)$} to evaluate an encoding/decoding scheme.
If we did not encode the information bit, the receiver has
$\Pr(\epsilon)=p$ in making a wrong decision. When we use repetitive
encoding, in the case that the channel corrupts first two bits,
different decoding schemes $\decode_{\ms{mv}}$ and
$\decode_{\ms{3rd}}$ have probabilities of failure
$\Pr(\epsilon)=p^2(1-p)$ and $\Pr(\epsilon)=0$, respectively.

From the example of repetition codes, we learn that
\begin{enumerate}
\item Encoding/decoding combination definitely helps error control.
The basic principle of encoding is to add ancillary bits (i.e.,
redundancy) to information messages; decoding is to find the
locations of errors, correct errors, and then remove ancillary bits.
\item A decoder able to correct errors depends on the error models
and decoding methods. A better decoding procedure can restore the
messages represented in the codewords after any errors occurred.
\end{enumerate}

\underline{\textbf{Linear Codes:}} The above example of repetition
codes encodes 1-bit information 0/1 into 3-bit codewords 000/111. In
fact, we can use matrix notations to represent 0/1 and 000/111. Let
column vectors $a_i=[i]$ and $c_i=[i,i,i]^T$, $i=0,1$, denote the
information bits and codewords, respectively. We can transform the
repetition encoding procedure as a matrix computation:
\begin{equation}
    c_i=\left[\begin{array}{c}1\\1\\1\end{array} \right]a_i \:,
\end{equation}
where all the arithmetic operations are done modulo 2.
The matrix $\left[\begin{array}{ccc}1&1&1\end{array} \right]^T$ is
called the \textbf{generator matrix} for the repetition code. The
generator matrix represents the rule how we encode information bits
to codewords. We can generalize this encoding process from
repetition codes to linear codes. A \textbf{linear code} encoding
$k$-bit messages into an $m$-bit code space is specified by an $m$
by $k$ generator matrix $G$ whose entries are all elements of
$\mathbb{Z}_2$, i.e., zeros and ones. We say that such a code is an
$[m,k]$ code. A slightly complicated example is to encode 2-bit
information into 4-bit codewords by duplicating each information
bit. Table~\ref{Tab:[4,2]code} tabulates this $[4,2]$ code, where
$(\cdot)$ are shorthand notations for the column vectors. The
generator matrix is
\begin{equation}
    G=\left[\begin{array}{cc}1&0\\1&0\\0&1\\0&1\end{array} \right]
    \:.
\end{equation}
 In general for large-sized linear codes, we only have
to specify generator matrices rather than specify all the
correspondence between information bits and codewords.

    \begin{table}[t]
    \caption{A $[4, 2]$ linear code.}
    \label{Tab:[4,2]code}
    \centering
    \begin{tabular}{|c|c|}
    \hline
    information bits & \makebox[2in]{codewords}\\
    \hline \hline
    $a_1=(0,0)$  & $c_1=(0,0,0,0)$ \\ \hline
    $a_2=(0,1)$  & $c_2=(0,0,1,1)$ \\ \hline
    $a_3=(1,0)$  & $c_3=(1,1,0,0)$ \\ \hline
    $a_4=(1,1)$  & $c_4=(1,1,1,1)$ \\ \hline
    \end{tabular}
    \end{table}

During the transmission of codewords through a binary symmetric
channel, noise operators do the following: $N_j$ maps codewords $c$
to $c'=c+n_j$, where $n_j$ is an $m$ by 1 unit vector with one at
$j$th entry and zeros elsewhere, and $+$ is bitwise addition modulo
2. For the case of $j=0$, $N_0$ is identity operator $I$
representing that no errors corrupt codewords, so $n_0=\zero$. To
decode, an $[m,k]$ code uses a \textbf{parity check matrix} $H$ with
size $m-k$ by $m$ such that $Hc=0$ for all the codewords $c$. Since
$c=Ga$, we have $HG=\zero$. Suppose that we want to decode $c$ but
we actually receive the corrupted version $c'$. It follows that
$Hc'=Hc+Hn_j=H n_j$. $H n_j$ is called the \textbf{error syndrome}
and is important in error correction.

The error syndrome provides cues of errors. Assume that there is no
error or only one error. In the case of no error, the error syndrome
is $\zero$. In the case of one error, the error syndrome is $H n_j$
telling us that the error occurs at the $j$th bit of the codeword.
Therefore, we can decode the corrupted codeword by flipping the
$j$th bit. However, if there are two errors, say $n_i$ and $n_j$,
the error syndrome becomes $H(n_i+n_j)$. If there is another error
$n_q$ such that $n_q=n_i+n_j$, ambiguity arises because we instead
will correct the $q$th bit. If all the errors $n_q \neq n_i+n_j$, we
either reject this codeword and then request the sender to
retransmit the codeword, or design another encoding/decoding scheme
that is able to immediately correct the codewords under two errors.

From the example of linear codes, we learn that
\begin{enumerate}

\item Error recovery essentially consists of two steps: error detection
and error correction. A receiver could have capability of error
detection alone. When the receiver detects errors, it requests the
sender to retransmit the codewords. On the other hand, the receiver
could be designed to correct errors on-site after detecting errors.
However, different decoding schemes have different capabilities to
correct errors.

\item Matrix representations of encoding/decoding procedure are
compact in code designs. The counterpart of matrices in quantum
mechanics is operators. Intuitively, we can use operators as
encoding and decoding operations in quantum error-correcting codes.

\item To correct
errors in codewords, we measure the syndrome that only contains the
information of errors. This concept is favorable in quantum
error-correcting codes. Measurement in quantum mechanics usually
collapses the target we attempt to measure. Measuring the syndrome
keeps the information of data intact and also tells how to correct
the errors.

\end{enumerate}

\subsection{Error Detection and Correction}\label{ssec:ClassicalErrorDetCorr}
In the above two examples, we saw that not all the errors are
correctable. In the example of linear codes, even though we have
detected an error, the error might be an ambiguous one when the
channel corrupts 2 bits. Therefore, we are inevitable to discuss the
detectability and correctability of errors.

\emph{Error detection} was used in the linear code example to reject
a codeword that could not be properly decoded. Error control methods
based on error detection alone work as follows: The receiver checks
whether the codeword is still in the code space $\code$; if yes, let
it go; if not, the result is rejected. The sender can be informed of
the failure so that the codeword can be resent. Given a set of noise
operators being protected against, the encoding/decoding scheme is
successful if for each noise operator, either the information is
unchanged, or the error is detected. Thus we can say that a noise
operator $N$ is \textbf{detectable} by a code if for each codeword
$c$ in the code, either $Nc = c$ or $Nc \not\in \code$. Of course,
the identity operator has no erroneous effects on codewords and is
always detectable. We can summarize the above observation in the
following theorem, see~\cite{Knill2002-IntroQuantECC}.

\begin{thm} \label{Thm:ClassDetect}
$N$ is detectable by a code if and only if
for all $c_m\ne c_n$ in the code, $Nc_m\ne c_n$.
\end{thm}

\emph{Error correction}, unlike error detection which is passive, is
active in the sense that decoder not only alarms errors but also
corrects errors as possible. Given a code $\code$ and a set $\noise$
of error operators $\set{N_0=\eye, N_1,N_2,\cdots}$, our goal is to
determine whether a decoding procedure exists such that $\noise$ is
correctable. Suppose that for some $c_m\ne c_n$ in the code and some
$i,j$, we have $c_q=N_ic_m=N_jc_n$. If, after an unknown error in
$\noise$ happened, the state $c_q$ is obtained, then it is not
possible to determine whether the original codeword was $c_m$ or
$c_n$, because we cannot tell whether $N_i$ or $N_j$ occurred. We
can formulate the correctability into the following theorem
\cite{Knill2002-IntroQuantECC}:

\begin{thm} \label{Thm:ClassCorrect}
$\noise$ is correctable by $\decode$ if and only if for all $c_m\ne
c_n$ in the code and for all $i,j$, it is true that $N_ic_m \ne
N_jc_n$.
\end{thm}

It is possible to relate the condition for correctability of an
error set to detectability. For simplicity, assume that each $N_i$
is invertible. The correctability condition is equivalent to the
statement that all products $N_j^{-1}N_i$ are detectable. To see the
equivalence, first suppose that some $N_j^{-1}N_i$ is not
detectable. Then there are $c_m\ne c_n$ in the code such that
$N_j^{-1}N_ic_m= c_n$. Consequently $N_ic_m=N_jc_n$ and the error
set is not correctable. Theorems~\ref{Thm:ClassDetect}
and~\ref{Thm:ClassCorrect} are observed from classical error
correction. They are also applicable to quantum error correction, as
we will see in Section~\ref{sec:QuantECC}.


\section{Quantum Error Correction} \label{sec:QuantECC}

Although classical coding theory has been developed to a
sophisticated level, people were not clear how to adopt the
classical ideas to quantum information until 1996, when
Shor~\cite{Shor1995-RedDecoherence} and
Steane~\cite{Steane1996-ECCinQuant} pointed out that quantum
error-correcting codes exist. Quantum coding theory has a difficulty
in that copying quantum information states is not possible. This is
known as the \textbf{no-cloning
theorem}~\cite{WootersZurek1982-NonCloning}. However, quantum error
correction works by circumventing this obstacle and demonstrates the
similarities to classical coding theory.

We start this section by introducing the units of quantum
information. Then we investigate error models of quantum channels.
Finally, we develop the quantum version of error correction.

\subsection{Quantum Bits} \label{ssec:QuBit}

The fundamental resource and basic unit of quantum information is
the \textbf{quantum bit}, coined as \textbf{qubit} by
Schumacher~\cite{Schumacher1995-QuantCoding}. A qubit behaves like a
classical bit enhanced by the superposition principle. From a
physical point of view, a qubit is represented by an ideal two-state
quantum system. Examples of such systems include photons (vertical
and horizontal polarization), electrons and other spin-$\frac{1}{2}$
systems (spin up and down), and atomic or ionic systems defined by
two energy levels.

From the information processing point of view, a qubit's state space
contains two logic states, or kets, $\ket{0}$ and $\ket{1}$. Their
Hermitian conjugates are denoted by bras $\bra{0}$ and $\bra{1}$.
The notation for these states was introduced by Dirac and is called
the \textbf{bra}-\textbf{ket} notation. Superpositions can be
expressed as sums $\alpha\ket{0}+\beta\ket{1}$ over the logical
states with complex coefficients. The complex numbers $\alpha$ and
$\beta$ are called the \textbf{amplitudes} of the superposition.
Such superpositions of distinguishable quantum states are one of the
basic tenets of quantum theory called the \textbf{superposition
principle}. Another way of writing a general superposition is as a
vector
\begin{equation} \label{eq:QubitMatrix}
    \alpha\ket{0}+\beta\ket{1} \leftrightarrow
    \left[\begin{array}{c}\alpha\\ \beta \end{array} \right]
\end{equation}
where the two-sided arrow $\leftrightarrow$ denotes the
correspondence between expressions that mean the same thing. It is
customary to assume that the vector has length 1, that is
$\magnitude{\alpha}^2+\magnitude{\beta}^2=1$.

\begin{figure*}[t]
    \setlength{\abovecaptionskip}{5pt}
    \centering
        \subfigure[Bits: states are either 0 or 1.]{
            \label{sFig:Bit}
            \begin{minipage}[b]{0.3\linewidth}
            \centering
            \setlength{\unitlength}{1.1cm}
            \begin{picture}(0.5,3.5)
                \thicklines
                \put(0,3.1){\circle*{0.2}}
                \put(0,0.8){\circle*{0.2}}
                \put(0.2,3){$0$}
                \put(0.2,0.7){$1$}
            \end{picture}
            \end{minipage}
        }
        \subfigure[Qubits: states are $\alpha\ket{0}+\beta\ket{1}$
                  with $\magnitude{\alpha}^2+\magnitude{\beta}^2=1$.]{
            \label{sFig:Qubit}
            \begin{minipage}[b]{0.5\linewidth}
            \centering
             \includegraphics[width=1.7in]{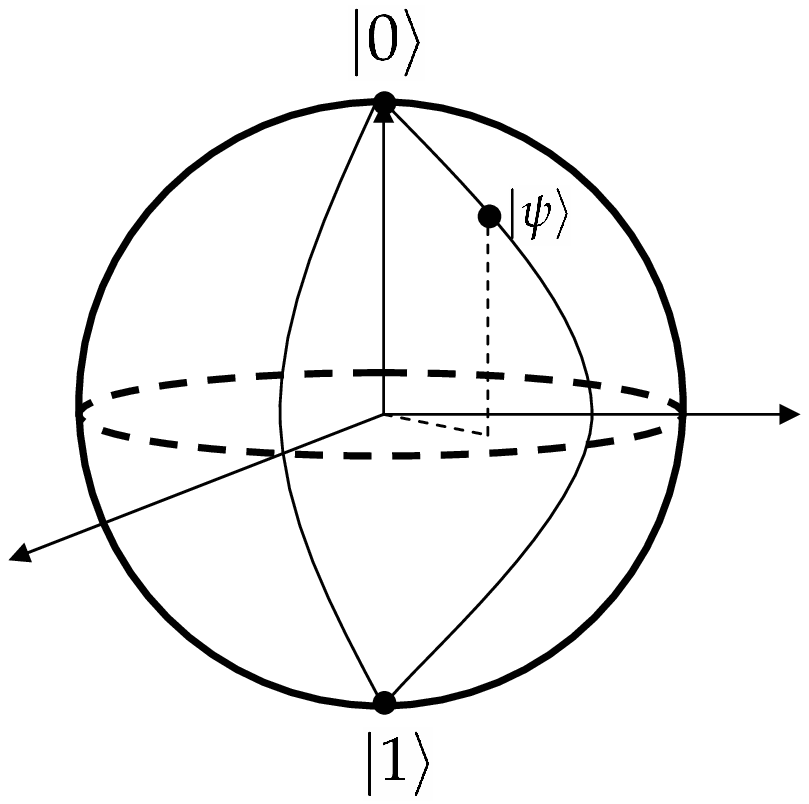}
            \end{minipage}
        }
    \caption{Visualization of bits versus qubits.}
    \label{Fig:BitQubit}
    \end{figure*}

What is the difference between bits and qubits? A visualization of
the difference between bits and qubits is shown in
Fig~\ref{Fig:BitQubit}. Apparently, qubits occupy a continuum of the
spherical space while bits only take two possible discrete points.
Bennett and Shor~\cite{BennettShor1998-QInfoTh} compare bits with
qubits in other respects and also list theirs roles in quantum
communications.

The quantum mechanical manipulations of qubits are carried out by
operators. For example, the $\NOT$ gate operates on
$\alpha\ket{0}+\beta\ket{1}$ to exchange the two logic states:
\begin{equation}
    \NOT(\alpha\ket{0}+\beta\ket{1})=\alpha\ket{1}+\beta\ket{0}.
\end{equation}
Similar to quantum states represented by vectors in
Eq~(\ref{eq:QubitMatrix}), we can represent operators by matrices.
In matrix representation, the $\NOT$ operator is equivalent to
$\left[\begin{array}{cc}0&1\\1&0\end{array} \right]$.

A qubit lives in a Hilbert space. The Hilbert space for several
qubits is the tensor product of Hilbert spaces for individual
qubits. The notation of tensor product is $\kron$. Similarly, an
aggregate of operators acting on the tensor product of qubits can be
represented by the tensor product of individual operators. For the
details about quantum mechanics see standard
textbooks~\cite{Griffiths2002-ConsistentQuantTh,
Liboff2003-QuantMech, Griffiths2005-QuantMech}.

Since qubits behave totally differently from classical bits, Nielsen
and Chuang~\cite{NielsenChuang2000-QCQI} summarize three
difficulties in quantum error correction:
\begin{enumerate}
\item \emph{Measurement destroys quantum information}: Observation
      in quantum mechanics generally destroys the quantum state
      under observation, and makes recovery impossible.
\item \emph{No cloning}: The no-cloning
      theorem~\cite{WootersZurek1982-NonCloning} states that there is
      no quantum operation taking a state $\ket{\psi}$ to
      $\ket{\psi}\kron\ket{\psi}$ for all states $\ket{\psi}$. In
      other words, we cannot design a repetition code by duplicating a
      state several times.
\item \emph{Errors are continuous}: Since a qubit is continuous,
      different errors on a single qubit form a continuum.
      We hence require infinite precision to determine which
      error occurred in order to correct it.
\end{enumerate}

To overcome the first difficulty, we have to recall lessons from
classical error-correcting codes. We would like to use the syndrome
in the decoding procedure. Measuring the syndrome alone will not
bother the information-carrying quantum states.

The second difficulty can be circumvented by embedding the
\emph{physical} basis $\set{\ket{0},\ket{1}}$ into a \emph{logical}
basis $\set{\ket{0_L}, \ket{1_L}}$ of code space:
\begin{eqnarray}
      \ket{0}&\rightarrow & \ket{0_L} \\
      \ket{1}&\rightarrow &\ket{1_L} \:.
\end{eqnarray}
That is, an encoder maps a state
$\ket{\psi}=\alpha\ket{0}+\beta\ket{1}$ to
$\alpha\ket{0_L}+\beta\ket{1_L}$. For example, we can encode a
message in the sense of a repetition code by
\begin{eqnarray}
    \ket{0_L}& \equiv & \ket{000}\\
    \ket{1_L}& \equiv & \ket{111} \:.
\end{eqnarray}
In the example of Shor code~\cite{Shor1995-RedDecoherence}, the
encoded basis is
\begin{eqnarray}
  &\ket{0_L}\equiv\frac{(\ket{000}+\ket{111})(\ket{000}+\ket{111})(\ket{000}+\ket{111})}{2\sqrt{2}}\\
  &\ket{1_L}\equiv\frac{(\ket{000}-\ket{111})(\ket{000}-\ket{111})(\ket{000}-\ket{111})}{2\sqrt{2}}
  \:.
\end{eqnarray}

The third challenge can be dealt with using the fact that any
operator on the space of one qubit can be written as a linear
combination of Pauli operators defined as:
\begin{eqnarray}
    &\eyeOp=\left[\begin{array}{cc}1&0\\0&1\end{array}\right], \quad
    \PauliX=\left[\begin{array}{cc}0&1\\1&0\end{array}\right],
    \nonumber \\
    &\PauliZ=\left[\begin{array}{cc}1&0\\0&-1\end{array}\right],\quad
    \PauliY=\left[\begin{array}{cc}0&-i\\i&0\end{array}\right]=i\PauliX\PauliZ
    \:.
\end{eqnarray}
These operators have effects on the qubit listed in
Table~\ref{Tab:PauliOps}~\cite{Gottesman2000-IntroQuantECC}. As long
as the decoder can correct errors of $\PauliX$, $\PauliY$, and
$\PauliZ$, it will correct any and all errors.

    \begin{table*}[t]
    \caption{The Pauli operators.}
    \label{Tab:PauliOps}
    \centering
    \begin{tabular}{|l||c|l|}
    \hline
    Identity & $\eyeOp=\left[\begin{array}{cc}1&0\\0&1\end{array}\right]$ & $\eyeOp\ket{a}=\ket{a}$\\
    \hline
    Bit Flip & $\PauliX=\left[\begin{array}{cc}0&1\\1&0\end{array}\right]$ & $\PauliX\ket{a}=\ket{a\oplus 1}$\\
    \hline
    Phase Flip & $\PauliZ=\left[\begin{array}{cc}1&0\\0&-1\end{array}\right]$ & $\PauliZ\ket{a}=(-1)^a\ket{a}$\\
    \hline
    Bit and Phase Flip $\quad$ & $\quad\PauliY=\left[\begin{array}{cc}0&-i\\i&0\end{array}\right]=i\PauliX\PauliZ\quad$
                      & $\PauliY\ket{a}=i(-1)^a\ket{a\oplus 1}\quad$\\
    \hline
    \end{tabular}
    \end{table*}

\subsection{Quantum Error Models} \label{ssec:QuantErrorModels}

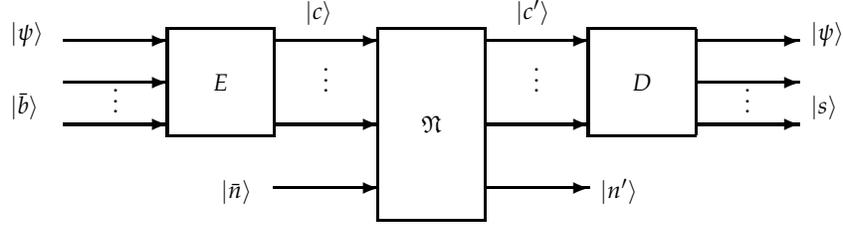
\begin{figure*}[t]
    \setlength{\abovecaptionskip}{-5pt}
    \centering
    \setlength{\unitlength}{1.4cm}
    \begin{picture}(9,2.5)
        \thicklines
        \put(0.5,1.9){$\ket{\psi}$}
        \put(0.5,1.2){$\ket{\bar{b}}$}
        \put(1,1.9){\vector(1,0){1}}
        \put(1,1.5){\vector(1,0){1}}
        \put(1,1.1){\vector(1,0){1}}
        \put(1,1.2){\makebox(1,0.4){\vdots}}
        \put(2,1){\framebox(1,1){$\Encode$}}
        \put(3.3,2.1){$\ket{c}$}
        \put(2.5,0.4){$\ket{\bar{n}}$}
        \put(3,1.9){\vector(1,0){1}}
        \put(3,1.1){\vector(1,0){1}}
        \put(3,1.4){\makebox(1,0.4){\vdots}}
        \put(3,0.5){\vector(1,0){1}}
        \put(4,0.2){\framebox(1,1.8){$\NoiseSet$}}
        \put(5,1.9){\vector(1,0){1}}
        \put(5,1.1){\vector(1,0){1}}
        \put(5,1.4){\makebox(1,0.4){\vdots}}
        \put(5,0.5){\vector(1,0){1}}
        \put(5.3,2.1){$\ket{c'}$}
        \put(6.1,0.4){$\ket{n'}$}
        \put(6,1){\framebox(1,1){$\Decode$}}
        \put(7,1.9){\vector(1,0){1}}
        \put(7,1.5){\vector(1,0){1}}
        \put(7,1.1){\vector(1,0){1}}
        \put(7,1.2){\makebox(1,0.4){\vdots}}
        \put(8.1,1.9){$\ket{\psi}$}
        \put(8.1,1.2){$\ket{s}$}
    \end{picture}
    \caption{Block diagram of a quantum communication system.}
    \label{Fig:QuantChannel}
    \end{figure*}

Error models in quantum communication are more complex than the
binary symmetric channel in classical communication. We now
introduce error models by starting with a model for a quantum
communication system, as shown in Fig~\ref{Fig:QuantChannel}. Let
the quantum information of interest be a ket $\ket{\psi}$ in a
Hilbert space $\Info$. Like adding the redundant bits in classical
coding, we consider ancillary qubits in quantum communication.
Ancillas are in the Hilbert space $\Ancilla$ and are initially in a
definite state $\ket{\bar{b}}$. Usually $\ket{\bar{b}}$ is set to be
$\ket{00\cdots0}$. In the encoding step, an encoder $\Encode$ is a
unitary transformation mapping $\Code=\Info\kron\Ancilla$ to itself.
That is, $\Encode(\ket{\psi}\kron\ket{\bar{b}})=\ket{c}\in \Code$.
$\Code$ is called \textbf{code space}.

With reference to Fig~\ref{Fig:QuantChannel}, errors in the channel
are induced by an interaction between $\Code$ and environment
$\Envi$ that has initial state $\ket{\bar{n}}$. The effect of this
interaction on $\Code$ is represented by a collection $\NoiseSet$ of
noise operators
$\set{\NoiseOp_0=\eyeOp,\NoiseOp_1,\NoiseOp_2,\cdots}$ mapping the
code $\Code$ to itself. These operators can always be chosen to be
linearly independent. The operator $\eyeOp$ in the collection
$\NoiseSet$ is an identity map. The effect of the interaction is
represented in the operator sum formalism
\begin{equation}
    \rho \rightarrow
    \NoiseSet(\rho)\define \sum_i \NoiseOp_i\rho \NoiseOp_i^{\dagger},
\end{equation}
where the normalization condition
\begin{equation}
    \sum_i \NoiseOp_i^{\dagger}\NoiseOp_i=\eyeOp
\end{equation}
ensures that $\trace{\NoiseSet(\rho)}=1$. There is an orthonormal
basis $\set{\ket{n_i}}$ on $\Envi$ such that for any $\ket{c}$ the
corrupted codeword is produced by a mapping
\begin{equation}
    \ket{c}\kron\ket{\bar{n}} \rightarrow \sum_i
    (\NoiseOp_i\ket{c})\kron\ket{n_i} \:.
\end{equation}

In the decoding step, a decoder $\Decode$ maps the noise corrupted
codeword $\ket{c'}$ back to $\ket{\psi}\kron\ket{s}$, where
$\ket{\psi}$ is the message we want to retrieve, and $\ket{s}$ is
the syndrome we will use to correct errors. Following the rationale
in classical error correction, we want to relate syndromes
$\ket{s_i}$ to errors $\NoiseOp_i$ homomorphically, and hope
$\ket{s_i}$ is independent of $\ket{\psi}$ for all $\ket{\psi}$ in
$\Info$. In short, we should design an encoding/decoding scheme to
reach a situation
\begin{equation}
    \Decode\NoiseOp_i\Encode(\ket{\psi}\kron\ket{\bar{b}})
    = \ket{\psi}\kron\ket{s_i} \:.
\end{equation}

Note that any operator can be written as a linear combination of
Pauli operators. In order to correct all kinds of errors, the
decoder should be able to correct errors of types $\PauliX$,
$\PauliY$, and $\PauliZ$, listed in Table~\ref{Tab:PauliOps}. The
quantum error models are apparently more complicated than the
classical error model of independent bit flips. To protect the
codewords against these noise operators, we next discuss the
conditions of error correctability.

\subsection{Conditions of Quantum Error Correction}

Similar to classical error correction, the procedure of quantum
decoding also consists of two steps: error detection and error
correction. In the detection step, we want to distinguish different
errors in the corrupted codewords. Then we apply the inverse of the
error operators in the correction step. If there are error operators
corrupting different codewords into an identical codeword, we face
ambiguity. In the sequel, the errors to be correctable must meet
some conditions.

The condition of quantum error correction is stated in the following
theorem~\cite{KnillLaflamme1997-TheoryQECC}:
\begin{thm} \label{Thm:QuantCorrect}
    Let $\CodeSub$ be a quantum code and $\ProjectOp$ be the projector onto
    $\CodeSub$. A set $\NoiseSet_{\ms{cr}}$ of noise operators $\set{\NoiseOp_i}$
    is correctable if and only if for all $i,j$
    \begin{equation}
        \label{eq:QuantCorrect}
        \ProjectOp \NoiseOp_i^\dagger \NoiseOp_j \ProjectOp = \lambda_{ij}\ProjectOp
    \end{equation}
    for some set of complex numbers $\set{\lambda_{ij}}$.
\end{thm}

Theorem~\ref{Thm:QuantCorrect} can be stated in terms of the
\emph{orthonormal basis} of $\CodeSub$. Let $\set{\ket{c_q}}$ be an
orthonormal basis of codewords which span the subspace $\CodeSub$.
Then we have
\begin{equation}
    P\ket{c_q}=\ket{c_q} \quad \mbox{and}\quad
    (\eyeOp-P)\ket{c_q}=0.
\end{equation}
Substituting this relation into Eq~(\ref{eq:QuantCorrect}) gives
\begin{eqnarray} \label{eq:QuantCorrect-UsingBasis}
    \bra{c_p}\NoiseOp_i^\dagger \NoiseOp_j \ket{c_q}
    = \lambda_{ij}\delta_{pq} \:,
\end{eqnarray}
which is an equivalent statement to Theorem~\ref{Thm:QuantCorrect}.
There are some insights for this theorem. We address them in the
following.

\begin{enumerate}
\item
\emph{Only some noise operators are correctable.} Recall a lesson in
classical error correction: \emph{not all the errors are
correctable, and the correctable errors depend on the decoding
scheme}. This means that we do not have an ambition to correct all
the error operators in $\NoiseSet$, but do focus only on a set
$\NoiseSet_{\ms{cr}}$ of correctable errors. Since
$\NoiseSet_{\ms{cr}}$ depends upon the decoding method $\Decode$, we
can relate them as $\NoiseSet_{\ms{cr}}(\Decode)$. Note that for a
given codeword space $\CodeSub$, there may be more than one possible
$\Decode$ and more than one possible family $\NoiseSet_{\ms{cr}}$ of
correctable errors. Conversely, given some decoding operation
$\Decode$, if $\set{\NoiseOp_i}$ is any collection of operators
drawn from $\NoiseSet_{\ms{cr}}(\Decode)$, then
Theorem~\ref{Thm:QuantCorrect} will be satisfied. When
Eq~(\ref{eq:QuantCorrect}) holds, the basis states $\set{\ket{c_q}}$
span a quantum error-correcting code.

\item
\emph{Linear space of noise operators.} The noise operators
$\set{\NoiseOp_i}$ in the set $\NoiseSet_{\ms{cr}}$ span a linear
space $\NoiseSpace$ of operators. We next would like to know whether
all the elements in $\NoiseSpace$ are correctable. If $U$ is a
matrix (not necessarily unitary) with entries $u_{mn}$, one can
define the new noise operators $F_j=\sum_i u_{ji}\NoiseOp_i $. The
left hand side of Eq~(\ref{eq:QuantCorrect}) becomes
\begin{eqnarray} \label{eq:QuantCorrect-BasisChange}
    PF_i^\dagger F_j P &=&
        P\left(\sum_{m,n} u_{im}^* \NoiseOp_m^\dagger u_{jn}\NoiseOp_n \right)P
    \nonumber\\
    &=& \sum_{m,n} u_{im}^* \lambda_{mn} u_{jn}  P
    \nonumber\\
    &=& (U^\dagger\Lambda U)_{ij} P \:,
\end{eqnarray}
where $\Lambda$ denotes the matrix $\lambda_{mn}$.
Equation~(\ref{eq:QuantCorrect-BasisChange}) shows that all the
elements in $\NoiseSpace$ are correctable. That is, if $\NoiseOp_i$
and $\NoiseOp_j$ are correctable, so is $\alpha \NoiseOp_i+ \beta
\NoiseOp_j$. Let
$\NoiseSet_{\ms{cr}}^{\ms{B}}=\set{\NoiseOp_1^{\ms{B}},\cdots,\NoiseOp_M^{\ms{B}}}$
be a basis of $\NoiseSpace$. It follows that
$\NoiseSet_{\ms{cr}}^{\ms{B}}$ satisfies
Eq~(\ref{eq:QuantCorrect-BasisChange}) with
$F_i=\NoiseOp_i^{\ms{B}}$.


\item
\emph{Principal errors.} There is a unitary transformation $U$ which
can diagonalize the Hermitian matrix $\Lambda$, i.e.,
\begin{equation} \label{eq:QuantCorrect-Diagonalize}
    \lambda_{ij}= \sum_k u_{ik}d_k u_{jk}^*
\end{equation}
with eigenvalues $d_k\ge0$. This is equivalent to defining a new set
of error operators $\set{F_k}$ such that
\begin{equation} \label{eq:QuantCorrect-DiagError}
    PF_k^\dagger F_l P = \delta_{kl} d_k P \:.
\end{equation}
For the case of $d_k>0$, we define \textbf{principal errors} $V_k$
as
\begin{equation} \label{eq:QuantCorrect-PincipalError}
    V_k \define \frac{1}{\sqrt{d_k}} F_k
\end{equation}
by normalizing $F_k$. It follows that $PV_k^\dagger
V_lP=\delta_{kl}P$, or equivalently
\begin{equation} \label{eq:QuantCorrect-PincipalErrorBasis}
    \bra{c_p}V_k^\dagger V_l\ket{c_q}= \delta_{kl}\delta_{pq} \:.
\end{equation}
Equation~(\ref{eq:QuantCorrect-PincipalErrorBasis}) is of
significance. We first denote $V_l\ket{c_q}$ as $\ket{c_q^l}$. When
$k\ne l$, Eq~(\ref{eq:QuantCorrect-PincipalErrorBasis}) reveals that
$\CodeSubK$ and $\CodeSubL$ are mutually orthogonal. When $k=l$,
Eq~(\ref{eq:QuantCorrect-PincipalErrorBasis}) becomes
$\braket{c_p^k}{c_q^k}=\delta_{pq}$. This means that $V_k$ unitarily
transforms the codeword space $\CodeSub$ spanned by
$\set{\ket{c_q}}$ onto another subspace $\CodeSubK$ spanned by
$\set{\ket{c_q^k}}$.

\item
\emph{Null errors.} In general, some of $d_k$ in
Eq~(\ref{eq:QuantCorrect-Diagonalize}) are zero. The \textbf{null
errors} are errors with $d_k=0$ such that
Eq~(\ref{eq:QuantCorrect-DiagError}) becomes $PF_k^\dagger F_k P
=0$, or, equivalently,
\begin{equation}
    \bra{c_q}F_k^\dagger F_k\ket{c_q}= 0 \:.
\end{equation}
Since $F_k$ are not zero, the role of $F_k$ is to annihilate
codewords and never contributes a component to the actual state of
the system. This simply means that these $F_k$ occur with zero
probability.

\end{enumerate}


\section{Fault-Tolerant Quantum Computation} \label{sec:FaultTole}

We have described quantum error correction. However, the
computations used during encoding and decoding are vulnerable to
errors. In the classical computer systems, the basic idea of
fault-tolerant computation is to add \emph{spatial} redundancy to
reduce the probability of failure. This concept applies to
fault-tolerant quantum computation as well.

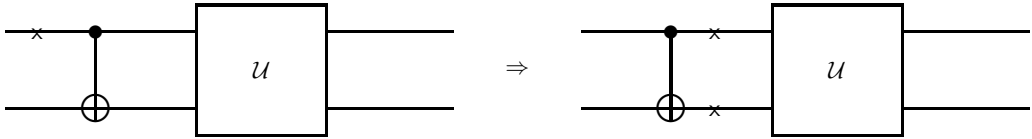
\begin{figure*}[t]
    \centering
    \setlength{\unitlength}{1.7cm}
      \newsavebox{\ErrProp}
      \savebox{\ErrProp}(0,0){
        \thicklines
        \put(0.5,0.8){\line(1,0){1.5}}
        \put(0.5,0.2){\line(1,0){1.5}}
        \put(1.2,0.8){\line(0,-1){0.7}}
        \put(1.2,0.8){\circle*{0.1}}
        \put(1.2,0.2){\circle{0.2}}
        \put(2,0){\framebox(1,1){$\mathcal{U}$}}
        \put(3,0.8){\line(1,0){1}}
        \put(3,0.2){\line(1,0){1}}
      }
    \begin{picture}(9,2.3)
      \put(0.7,0.74){$\mathsf{x}$}
      \put(0,0.5){\usebox{\ErrProp}}
      \put(4,0){\makebox(1,1){$\Rightarrow$}}
      \put(4.5,0.5){\usebox{\ErrProp}}
      \put(6,0.74){$\mathsf{x}$}
      \put(6,0.14){$\mathsf{x}$}
    \end{picture}
    \caption{An error (denoted by $\mathsf{x}$) at the top input of
             $\CNOT$ gate propagating through the $\CNOT$
             gate results in that $U$ operates on the wrong inputs.}
    \label{Fig:ErrorPropagation}
    \end{figure*}

Fig.~\ref{Fig:ErrorPropagation} is an example of Controlled-NOT
($\CNOT$) gate followed by an operation $U$. Unfortunately, there is
an error (denoted by $\mathsf{x}$) at one of the inputs of $\CNOT$
gate. The propagation of this error through $\CNOT$ gate generates a
catastrophe because $U$ operates on the wrong inputs. To avoid
spreading errors through a quantum circuit, we have to scrutinize
the behavior of error propagation and design fault-tolerant quantum
computation in a systematic way.

\subsection{The Laws of Fault-Tolerant Quantum Computation}
Preskill~\cite{Preskill1998-ReliableQC,Preskill1998-FTQC} studied
quantum error behavior and distilled five laws to design
fault-tolerant quantum computation. We summarize them in this
section.

\begin{enumerate}
\item \emph{Do not use the same qubit twice}. We use an example in
      Fig.~\ref{sFig:NotSameBitTwice-Bad} to illustrate this law. In
      Fig.~\ref{sFig:NotSameBitTwice-Bad}, the data consists of
      two qubits and the ancilla is one qubit. If there is an error
      in the ancilla, this error will spread catastrophically to the entire
      circuit. Note that quantum errors can be a bit flip, a phase flip, or
      a combination. A bit flip error in a $\CNOT$ circuit propagates
      from the source to the target. A phase error, however, goes in the opposite
      direction, from the target to the source. The improvement of this
      circuit is to decompose the ancilla into two qubits, as shown
      in Fig.~\ref{sFig:NotSameBitTwice-Good}. The new design
      guarantees that an error in one of the ancilla qubits only affects the circuit once.

    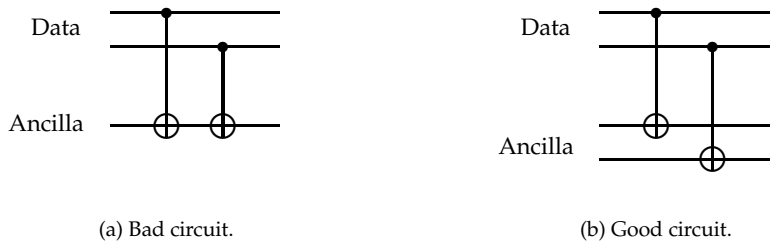
\begin{figure*}[t]
    \centering
        \subfigure[Bad circuit.]{
            \label{sFig:NotSameBitTwice-Bad}
            \begin{minipage}[b]{0.35\linewidth}
            \centering
            \setlength{\unitlength}{1.5cm}
            \begin{picture}(2,1.6)
            \thicklines
            \put(-0.2,1.3){Data}
            \put(-0.4,0.45){Ancilla}
            \put(0.5,1.5){\line(1,0){1.5}}
            \put(0.5,1.2){\line(1,0){1.5}}
            \put(0.5,0.5){\line(1,0){1.5}}
            \put(1,1.5){\line(0,-1){1.1}}
            \put(1,1.5){\circle*{0.1}}
            \put(1,0.5){\circle{0.2}}
            \put(1.5,1.2){\line(0,-1){0.8}}
            \put(1.5,1.2){\circle*{0.1}}
            \put(1.5,0.5){\circle{0.2}}
            \end{picture}
            \end{minipage}
        }
        \subfigure[Good circuit.]{
            \label{sFig:NotSameBitTwice-Good}
            \begin{minipage}[b]{0.35\linewidth}
            \centering
            \setlength{\unitlength}{1.5cm}
            \begin{picture}(2,1.3)
            \thicklines
            \put(-0.2,1.3){Data}
            \put(-0.4,0.25){Ancilla}
            \put(0.5,1.5){\line(1,0){1.5}}
            \put(0.5,1.2){\line(1,0){1.5}}
            \put(0.5,0.5){\line(1,0){1.5}}
            \put(0.5,0.2){\line(1,0){1.5}}
            \put(1,1.5){\line(0,-1){1.1}}
            \put(1,1.5){\circle*{0.1}}
            \put(1,0.5){\circle{0.2}}
            \put(1.5,1.2){\line(0,-1){1.1}}
            \put(1.5,1.2){\circle*{0.1}}
            \put(1.5,0.2){\circle{0.2}}
            \end{picture}
            \end{minipage}
        }
    \caption{Do not use the same bit twice.}
    \label{Fig:NotSameBitTwice}
    \end{figure*}

\item \emph{Copy the errors, not the data qubits}.
      We want to copy onto the ancilla the information about the
      errors in the data qubits, without inducing additional errors
      into the data. To achieve this goal, we
      must prepare an appropriate state of the ancilla before we copy any
      information. This ancillary state is carefully chosen so that by measuring the ancilla we
      acquire only information about the errors having occurred,
      and don't perturb the encoded data.

\item \emph{Verify when we encode a known quantum state}.
      The encoding process is vulnerable to errors---the power
      of the code to protect against channel noises is not yet in place.
      A single error may propagate virulently, as we saw in
      Fig.~\ref{Fig:NotSameBitTwice}. Therefore, we
      should carry out a measurement which checks that the encoding has
      been done correctly.

\item \emph{Repeat operations}.
      Operations, such as encoding verification and
      syndrome measurement, themselves may be erroneous. For instance,
      errors while measuring a syndrome can both damage the data and generate an erroneous
      syndrome. Thus we have to repeat an operation to increase
      our confidence that the operation was performed correctly.

\item \emph{Use the right codes}. The code we use for computation
      should have special properties so that we can apply quantum
      gates to the encoded information that operate efficiently and
      that adhere to the preceding four laws. For example, a good
      code for computation might be such that a $\CNOT$ gate acting on
      encoded qubits is implemented as in Fig.~\ref{Fig:BitwiseCNOT}
      ---with a single $\CNOT$ gate applied to each bit in both the
      source block and the target block.

    \begin{figure*}[t]
    \setlength{\abovecaptionskip}{0pt}
    \centering
        \setlength{\unitlength}{1.5cm}
        \begin{picture}(5.5,3)
        \thicklines
        \put(0,1.2){$\ket{\psi}$}
        \put(0,0.25){$\ket{b}$}
        \put(0.5,1.3){\line(1,0){1.2}}
        \put(0.5,0.3){\line(1,0){1.2}}
        \put(1.1,1.3){\line(0,-1){1.1}}
        \put(1.1,1.3){\circle*{0.1}}
        \put(1.1,0.3){\circle{0.2}}
        \put(2,0.5){\makebox(0.5,0.5){{\Large =}}}
        \put(2.8,1.2){$\ket{\psi}$}
        \put(2.8,0.25){$\ket{b}$}
        \put(3.2,1.5){\line(1,0){2}}
        \put(3.2,1.3){\line(1,0){2}}
        \put(3.2,1.1){\line(1,0){2}}
        \put(3.2,0.5){\line(1,0){2}}
        \put(3.2,0.3){\line(1,0){2}}
        \put(3.2,0.1){\line(1,0){2}}
        \put(3.8,1.5){\line(0,-1){1.1}}
        \put(3.8,1.5){\circle*{0.1}}
        \put(3.8,0.5){\circle{0.2}}
        \put(4.1,1.3){\line(0,-1){1.1}}
        \put(4.1,1.3){\circle*{0.1}}
        \put(4.1,0.3){\circle{0.2}}
        \put(4.4,1.1){\line(0,-1){1.1}}
        \put(4.4,1.1){\circle*{0.1}}
        \put(4.4,0.1){\circle{0.2}}
        \end{picture}
    \caption{Bitwise implementation of a $\CNOT$ gate.}
    \label{Fig:BitwiseCNOT}
    \end{figure*}
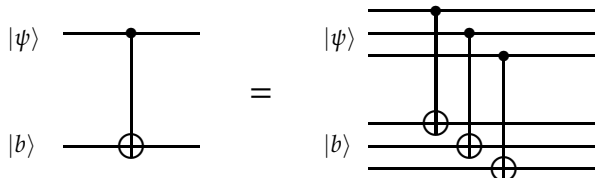

\end{enumerate}

\subsection{Concatenated Codes}

These five laws provide a guideline to ensure fault-tolerant quantum
computation. In short, we spend spatial resources to reduce failure
rate. Figs.~\ref{Fig:NotSameBitTwice} and~\ref{Fig:BitwiseCNOT} have
demonstrated this concept. Besides making quantum computation
reliable, another goal of fault-tolerant quantum computation is to
build a scalable quantum information processor. The \emph{accuracy
threshold theorem} shows how to achieve these goals. The accuracy
threshold theorem~\cite{Knill1996-ThreAccuracy} states the
following:
\begin{thm} \label{Thm:AccuracyThreshold}
If the error rate per qubit is less than a threshold, then an
arbitrarily long quantum computation can be executed with high
reliability.
\end{thm}

\begin{figure*}[tbp]
        \setlength{\abovecaptionskip}{0pt}
        \centering
        \includegraphics[width=4.5in]{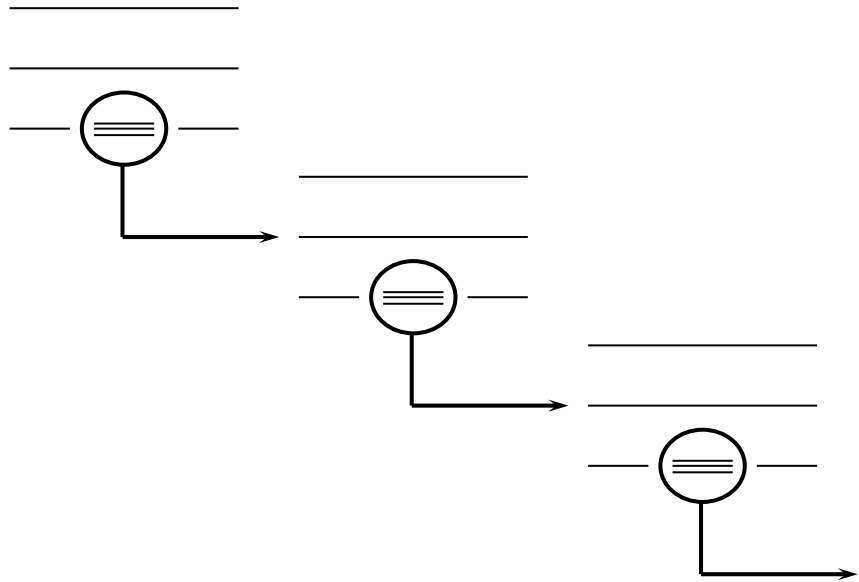}
        \caption{The idea of concatenated codes.}
        \label{Fig:ConcatenatedCode}
    \end{figure*}

A feasible realization of Theorem~\ref{Thm:AccuracyThreshold} is
\emph{concatenated codes}~\cite{Knill1996-ConcaQuantCode}. A
concatenated code is obtained by repeating the following
construction several times until a tolerated error rate is achieved.
The procedure for constructing a concatenated code is in the
following. Suppose we have an error-correcting code $\mathcal{C}$
with size $[m,1]$; i.e., encoding one information qubit into $m$
qubits. In fact, if we magnify these $m$ qubits, every qubit is not
really a single qubit but another block of $m$ qubits acquired by
encoding this qubit via the same code $\mathcal{C}$. In other words,
we actually encode the information qubit into $m^2$ qubits. If we
again use the code $\mathcal{C}$ to encode each qubit in the second
layer to obtain the third level, we essentially encode the initial
information qubit into $m^3$ qubits. If there are $L$ levels, the
information qubit is encoded in a block of $m^L$ qubits. This
procedure is called \textbf{concatenation}, illustrated in
Fig.~\ref{Fig:ConcatenatedCode}. Note that we have spent more
spatial resources (circuit area) to encode one information qubit.
However, we don't increase the complexity of the code $\mathcal{C}$,
because, no matter in which layer the code is used, we just build up
a hierarchial coding architecture by systematically coping many
times the fundamental circuit of $\mathcal{C}$.

Next, we have to study the performance of using concatenation, to
ensure the method provides a better protection. Suppose that the
errors are independent events and the probability of error per qubit
is $p$. If the code $\mathcal{C}$ is able to correct errors in $g$
of the $m$ qubits, the total probability of recovery failure is
\begin{equation} \label{eq:FT-ProbError}
    \Prob_1(\epsilon)=\sum_{i=g+1}^m \gamma_i p^i \:,
\end{equation}
where the coefficient $\gamma_i$ captures the combinatorial effects
for the occurrence of $i$ errors. If $p$ is small enough,
$\Prob_1(\epsilon)$ can be bounded as
\begin{equation}
    \Prob_1(\epsilon)\le \Gamma p^{g+1}
\end{equation}
by introducing a constant $\Gamma$. Now we consider a concatenation
code with two levels. Recall that there are $m$ sub-blocks of size
$m$ qubits. This two level coding architecture fails to correct
errors only when there are uncorrectable errors (more than $g$
errors) in more than $g$ sub-blocks. In this case, the probability
of failure is
\begin{equation}
    \Prob_2(\epsilon)\le (\Gamma p^{g+1})^{g+1} \:.
\end{equation}

If we use an $L$-level concatenation code, the failure rate reduces
to
\begin{equation} \label{eq:FT-L levelErrorRate}
    \Prob_L(\epsilon)\le \Gamma^{(g+1)^{L-1} } p^{(g+1)^L} \:.
\end{equation}
Note that Eq~(\ref{eq:FT-L levelErrorRate}) is an \emph{$L$-fold
exponentially decreasing} function of $L$. This explains why the
concatenation is efficient in reducing the failure rate.

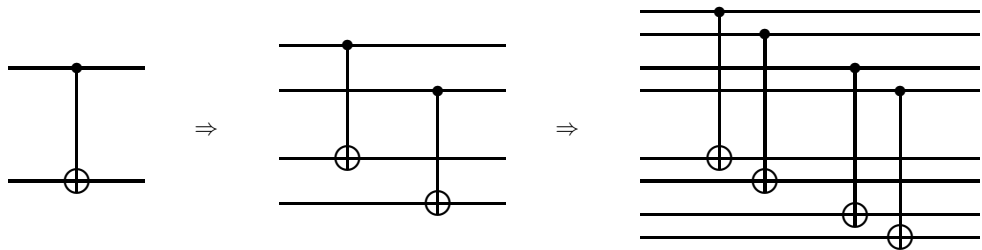
\begin{figure*}[tbp]
    \centering
        \setlength{\unitlength}{1.5cm}
        \begin{picture}(9.5,2.5)
        \thicklines
        \put(0.5,1.3){\line(1,0){1.2}}
        \put(0.5,0.3){\line(1,0){1.2}}
        \put(1.1,1.3){\line(0,-1){1.1}}
        \put(1.1,1.3){\circle*{0.1}}
        \put(1.1,0.3){\circle{0.2}}
        \put(2,0.5){\makebox(0.5,0.5){$\Rightarrow$}}
        \put(2.9,1.5){\line(1,0){2}}
        \put(2.9,1.1){\line(1,0){2}}
        \put(2.9,0.5){\line(1,0){2}}
        \put(2.9,0.1){\line(1,0){2}}
        \put(3.5,1.5){\line(0,-1){1.1}}
        \put(3.5,1.5){\circle*{0.1}}
        \put(3.5,0.5){\circle{0.2}}
        \put(4.3,1.1){\line(0,-1){1.1}}
        \put(4.3,1.1){\circle*{0.1}}
        \put(4.3,0.1){\circle{0.2}}
        \put(5.2,0.5){\makebox(0.5,0.5){$\Rightarrow$}}
        \put(6.1,1.8){\line(1,0){3}}
        \put(6.1,1.6){\line(1,0){3}}
        \put(6.1,1.3){\line(1,0){3}}
        \put(6.1,1.1){\line(1,0){3}}
        \put(6.1,0.5){\line(1,0){3}}
        \put(6.1,0.3){\line(1,0){3}}
        \put(6.1,0){\line(1,0){3}}
        \put(6.1,-0.2){\line(1,0){3}}
        \put(6.8,1.8){\line(0,-1){1.4}}
        \put(6.8,1.8){\circle*{0.1}}
        \put(6.8,0.5){\circle{0.2}}
        \put(7.2,1.6){\line(0,-1){1.4}}
        \put(7.2,1.6){\circle*{0.1}}
        \put(7.2,0.3){\circle{0.2}}
        \put(8,1.3){\line(0,-1){1.4}}
        \put(8,1.3){\circle*{0.1}}
        \put(8,0){\circle{0.2}}
        \put(8.4,1.1){\line(0,-1){1.4}}
        \put(8.4,1.1){\circle*{0.1}}
        \put(8.4,-0.2){\circle{0.2}}
        \end{picture}
    \caption{$\CNOT$ gate with three levels of concatenation.}
    \label{Fig:ConcatenatedCode-2Qubits}
    \end{figure*}

Finally, we use a simple example to illustrate the concatenated
codes. Fig.~\ref{Fig:ConcatenatedCode-2Qubits} is a case of $\CNOT$
gate with three levels of concatenation. In the first level, we have
a generic $\CNOT$ gate. In the second level, each wire is decomposed
into two wires; the first-level $\CNOT$ gate now is implemented by
two pairs of sub-$\CNOT$ gates. To obtain the third level, each wire
in the second level is split into two wires. Now the four pairs of
sub-$\CNOT$ gates aggregately perform as a single $\CNOT$ gate in
the first level. By creating one more level, we can reduce the
probability of failure significantly. In fact, there is no free
lunch for reducing failure rate, because we already increase the
circuit area and complicate the circuit, as
Fig.~\ref{Fig:ConcatenatedCode-2Qubits} shows.

\section{Conclusions} \label{sec:Conclusions}

This report investigates the fundamentals of quantum
error-correcting codes. The difference between quantum and classical
communication systems results from the different characteristics of
bits and qubits and from different error models for the noisy
channels. Nevertheless, quantum error correction shares many
concepts with its classical counterpart. Both quantum and classical
coding schemes add ancillary qubits/bits and measure a syndrome to
protect information messages. They also have similar conditions for
errror detectability and correctability. Due to lack of time, this
report leaves out the construction of quantum error-correcting
codes, which is now based on Gottesman's stabilizer
codes~\cite{Gottesman1997-PhD}.

The research in quantum error-correcting codes has already migrated
to nonbinary codes~\cite{Rains1999-Nonbinary}. However, there is
still a long ways to go. In order to use ancillary bits efficiently,
modern classical coding theory already goes beyond linear codes. The
codes frequently used in practice are nonlinear codes, such as tree
codes, trellis codes, turbo codes, and low-density parity-check
codes. The nonlinear version of quantum codes is still waiting
exploration.

In addition, the typical size of classical codes in usage is
$[m=2^{40},k=2^{20}]$, which is unreachable by contemporary quantum
codes and fault-tolerant quantum computation. To design a scalable
system of quantum information processing, we expect more innovation
in the future. Quantum error correction is still in its toddler
stage, but we believe that the 21st century will be the golden age
of quantum error correction.


\bibliography{Ref-JournalAbbr,Ref-QuantPhysInfo,Ref-InfoTheoCoding}
\bibliographystyle{IEEEtranS}


\end{document}